\documentclass[%
 aip,apl
 amsmath,amssymb,
 reprint,%
onecolumn
]{revtex4-1}

\usepackage{graphicx}
\usepackage{epstopdf}

\usepackage{dcolumn}
\usepackage{bm}
\usepackage{amsmath}
\usepackage{color} 
\usepackage{siunitx}
\usepackage[mathlines]{lineno}
\usepackage{hyperref}
\usepackage{float}
\usepackage[capitalize]{cleveref}

\begin{document}


\title{Robust nano-fabrication of an integrated platform for spin control in a tunable microcavity}

\author{Stefan Bogdanovi\'{c}}
\thanks{S. Bogdanovi\'{c} and M.S.Z. Liddy contributed equally to this work.}
\affiliation{QuTech, Delft University of Technology, PO Box 5046, 2600 GA Delft, The Netherlands}
\author{Madelaine S.Z. Liddy}
\thanks{S. Bogdanovi\'{c} and M.S.Z. Liddy contributed equally to this work.}
\affiliation{Institute for Quantum Computing, Department of Electrical and Computer Engineering, University of Waterloo, Waterloo, ON, N2L 3G1, Canada}
\author{Suzanne B. van Dam}%
\affiliation{QuTech, Delft University of Technology, PO Box 5046, 2600 GA Delft, The Netherlands}
\author{Lisanne C. Coenen}
\affiliation{QuTech, Delft University of Technology, PO Box 5046, 2600 GA Delft, The Netherlands}
\author{Thomas Fink}
\affiliation{Institute of Quantum Electronics, ETH Z\"urich, CH-8093 Z\"urich, Switzerland}
\author{Marko Lon\v{c}ar}
\affiliation{John A. Paulson School of Engineering and Applied Sciences, Harvard University, Cambridge, Massachusetts 02138, USA}
\author{Ronald Hanson}
\thanks{Electronic mail: r.hanson@tudelft.nl}
\affiliation{QuTech, Delft University of Technology, PO Box 5046, 2600 GA Delft, The Netherlands}

\date{\today}

\begin{abstract}
Coupling nitrogen-vacancy centers in diamond to optical cavities is a promising way to enhance the efficiency of diamond based quantum networks. An essential aspect of the full toolbox required for the operation of these networks is the ability to achieve microwave control of the electron spin associated with this defect within the cavity framework. Here, we report on the fabrication of an integrated platform for microwave control of an NV center electron spin in an open, tunable Fabry-P\'{e}rot microcavity. A critical aspect of the measurements of the cavity's finesse reveals that the presented fabrication process does not compromise its optical properties. We provide a method to incorporate a thin diamond slab into the cavity architecture and demonstrate control of the NV center spin. These results show the promise of this design for future cavity-enhanced NV center spin-photon entanglement experiments.\\
\end{abstract}

\maketitle

Nitrogen-vacancy (NV) colour centers in diamond have emerged as attractive candidates for quantum photonic applications. Their electronic spin can be optically initialized, read out in a single shot \citep{Robledo2011}, and coherently manipulated with the use of microwave signals\citep{Jelezko2004}. This spin-photon interface provides a platform for distant entanglement generation \citep{Hannes2013}, while additional coupling to nearby carbon-13 nuclear spins forms a multi-qubit quantum node\citep{Dutt2007,Neumann2010,Taminiau2012,Reiserer2016}. These aspects make the NV center a good candidate for quantum network protocols\citep{Kimble2008,NorbertPeter2017,Suzanne2017}.
The efficiency of entanglement generation between network nodes is currently limited by the NV center's low ($\approx 3 \%$) emission rate of the resonant zero-phonon line (ZPL) photons. This problem can be addressed by coupling NV centers to optical microcavities\cite{Benson,Barclay2011,Toeno_phc,Faraon_phc,Kaupp,Becher,Marko_phc,Hu_phc,Becher_phc,Englund2010,Englund_phc,JasonSmith,Kaupp_arxiv}, enhancing the ZPL emission rate and providing efficient photon extraction by means of the Purcell effect\citep{Purcell}. An appealing cavity design consists of an open, tunable Fabry-P\'{e}rot microcavity housing a large area diamond membrane\citep{Lily,Bogdanovic,Riedel} in which emitters retain their bulk-like properties\citep{Chu}. The tunability of this design enables both spectral positioning of the cavity to be resonant with the emitter as well as selective lateral placement of the emitter within the center of the cavity mode. However, in order to use these emitters in quantum information protocols, microwave control must be integrated into the cavity architecture. Here, we present fabrication methods used to create a platform that integrates microwave control of an NV center spin within an optical cavity while maintaining the cavity's high finesse properties. While microwave addressing of a single NV center spin has already been realized in thin diamond slabs\citep{Hodges2012} and photonic crystal cavities\citep{Englund_phc}, this is the first demonstration of NV center spin addressing within a framework tailored to the implementation of a tunable microcavity.

\begin{figure}
   \includegraphics[width=1\textwidth]{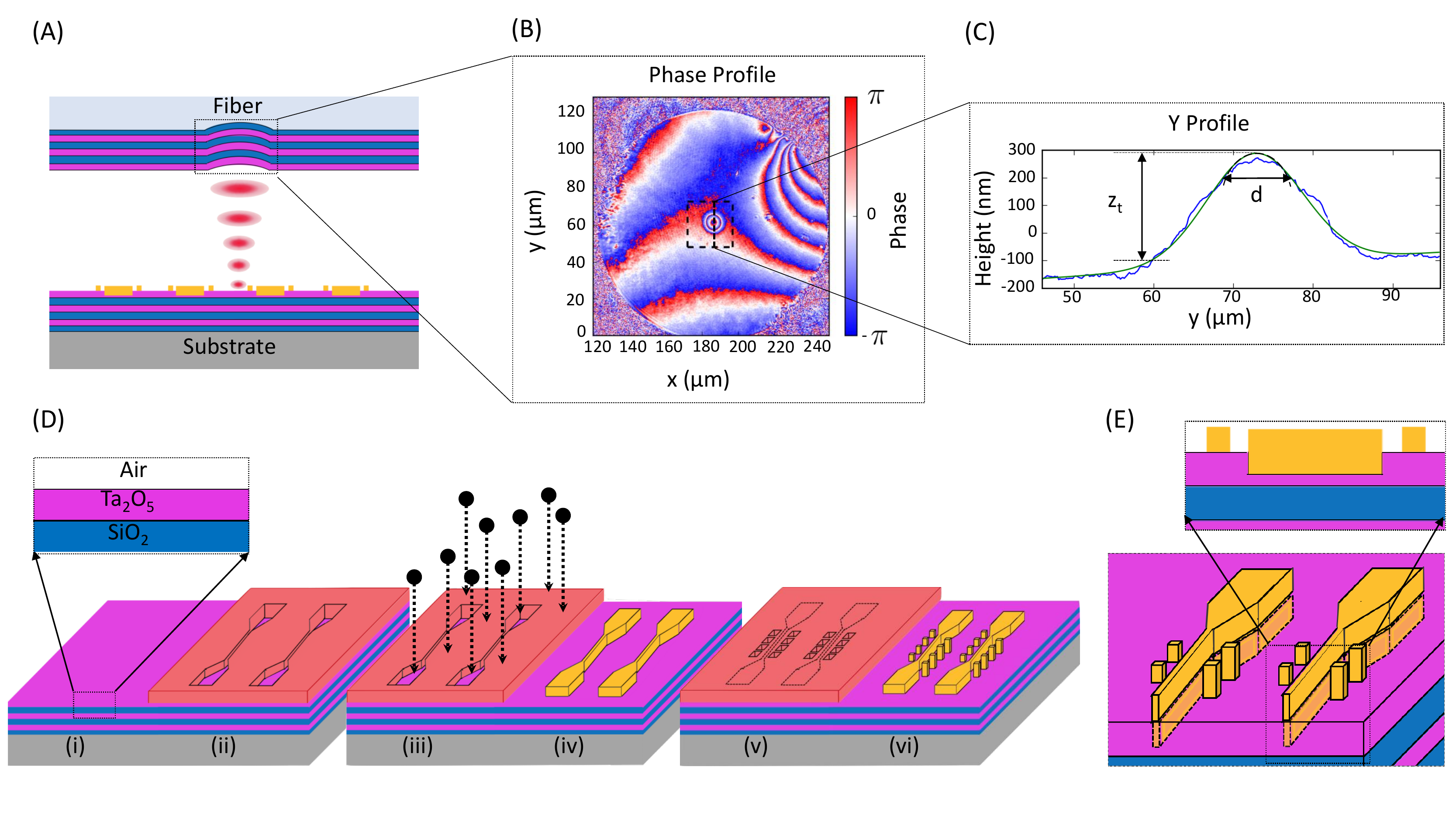}
	\caption{(Colour Online) Fabrication of the Fabry-P\'{e}rot cavity components. (A) The design of the Fabry-P\'{e}rot cavity, comprised of a processed planar mirror and laser ablated fiber interface. (B) Phase profile of the ablated fiber facet measured using an interferometer. (C) Reconstructed surface profile of the region denoted by the black rectangle in (B) (blue line) and fit with a Gaussian function (green line). The extracted parameters are $d = 8.3$ \SI{}{\micro m} and $z{_t} = 403$ nm. The fiber radius of curvature $R$, is 22.4 \SI{}{\micro m}. (D) Planar mirror processing stages: (i) Mirror coatings consist of 31 layers of alternating high (Ta$_{2}$O${_5}$, 75 nm thickness) and low (SiO${_2}$, 100 nm thickness) refractive index materials deposited onto a low roughness fused silica plate and ablated fiber tip. (ii) Patterned and developed optical photoresist mask for the microwave striplines. (iii) Etching 65 nm of material into the mirror stack using an ${SF_6}$ and ${O_2}$ ICP RIE recipe followed by (iv) evaporation of 5 nm (65 nm) of titanium (gold) and an overnight liftoff. (v) Patterned and developed optical resist mask for marker array, aligned to the gold striplines. (vi) Evaporation of 5 nm (5 nm) titanium (gold) followed by one hour of liftoff. (E) Cross-section of the mirror post processing revealing the embedded microwave striplines and surface marker array.}
   \label{fig:Fig_1}
\end{figure}

The cavity consists of a dimpled fiber tip and polished fused silica plate, both coated with a highly reflective dielectric mirror stack (\Cref{fig:Fig_1}(A)). Microwave striplines and marker arrays are fabricated on the planar mirror surface in order to locate the NV centers and address their spin within the diamond slab bonded to the mirror.\\

\paragraph{Fiber dimple:}
The curved fiber profile was fabricated using a CO$_2$ laser ablation technique\citep{Hunger2010,Fiber_hunger}: A single 1 ms long circularly polarized laser pulse is focused onto the cleaved fiber facet. As a result of thermal evaporation and subsequent melting, a concave depression with low surface roughness of $\sigma_{\textrm{rms}} \lesssim 0.20\pm0.02$\,nm is created. The depth and diameter of this depression can be controlled by varying a combination of the pulse power, duration, and beam waist. Due to the small fiber core diameter, care must be taken to center the depression onto the cleaved facet to ensure a good coupling efficiency to the cavity mode. Following dimple creation, the surface geometry is measured, \emph{in situ}, with an interferometer. \Cref{fig:Fig_1}(B) shows an exemplary phase measurement and the corresponding reconstructed surface profile (\Cref{fig:Fig_1}(C)). The shape of the depression closely resembles that of a two dimensional Gaussian. In its center, it can be approximated by a hemisphere with radius of curvature, $R \approx d^2/\left( 8 z_t \right)$, where $d$ and $z_t$ are the diameter and depth of the dimple, respectively\citep{Hunger2010}. Low-ellipticity profiles with comparable radii of curvature along the $x$- and $y$-direction are required in order to minimize polarization splitting of the cavity mode\citep{Uphoff2014}. The extracted ellipticity of the fiber used in this experiment is 1.3 \%.

\paragraph{Mirror coatings:}
Following laser ablation of the fibers, a mirror stack forming a Distributed Bragg Reflector (DBR) is deposited onto both the fiber facets and polished fused silica plates (Laseroptik). The residual transmission of this stack is measured to be 50 ppm at 637 nm wavelength. The observed fiber and specified fused silica surface roughness (0.2 nm and 0.5 nm RMS, respectively) correspond to scattering losses of 25 and 100 ppm \citep{Scattering}. Following coating, the planar mirror is annealed in vacuum at ${300^\circ}$C for 5 hours, which reduces the absorption losses of the stack from $\approx$ 50 ppm to $\leq$ 10 ppm\citep{Laseroptik}. The total losses give an expected value of the cavity finesse of $F \approx$ 22 000.

\paragraph{Striplines and marker field:}
In order to address the spin and index the location of the NV centers, parallel microwave striplines and a field of uniquely identifiable markers are fabricated on the planar mirror surface. For successful diamond bonding, the processed mirrors must possess a low profile for all patterned surface features. A two-step fabrication procedure was devised, which includes first patterning embedded microwave striplines into the planar mirror, followed by the deposition of a small marker array on the surface. The fabrication procedure is presented in \Cref{fig:Fig_1}(D). To begin the processing, optical photoresist, AZ 3007, is deposited on the planar mirror at a thickness of 1 \SI{}{\micro m}, followed by a soft bake at ${90^\circ}$C for 60 s. 4 mm long by 45 \SI{}{\micro m} wide masks for the striplines are patterned with a laser writer (DMO Microwriter ML-2), and developed in MF-321 for 45 s (\Cref{fig:Fig_1}(D,i-ii). Using an SF${_6}$ and O${_2}$ based Inductively Coupled Plasma Reactive Ion Etch (ICP RIE) to remove 65 nm of material, the mask for the microwave striplines is transferred into the planar mirror (\Cref{fig:Fig_1}(D,iii)) (Methods). After etching, a 5 nm layer of titanium is evaporated as an adhesion layer followed by 65 nm of gold (Temescal FC2000). The first stage of the fabrication is completed by removing all excess gold with an overnight liftoff in photoresist stripper (PRS) at ${70^\circ}$C, leading to the result shown in (\Cref{fig:Fig_1}(D,iv)). The second stage of the fabrication procedure proceeds with the same photoresist deposition method, followed by patterning the marker array, aligned adjacent to the microwave striplines, and then developed (\Cref{fig:Fig_1}(D,v)). 5 nm layers of Ti and Au are evaporated consecutively, followed by 1 hour of liftoff in PRS at ${70^\circ}$C to complete the fabrication (\Cref{fig:Fig_1}(D,vi)).

Essential to the desired cavity architecture is the possibility to bond a several micrometer thin diamond membrane over the structures on the planar mirror. Irregular structures and frills on the edges of the fabricated striplines have been found to deter successful bonding. Furthermore, a microwave stripline that is fully recessed in the etched trench acts as a capillary channel for water used during the bonding process, preventing a good diamond-mirror bond. A uniform raised profile of a feature above the mirror surface by no more than 20 nm was found to allow for successful bonding.

\paragraph{Cavity finesse:}

\begin{figure}
   \includegraphics[width=\textwidth]{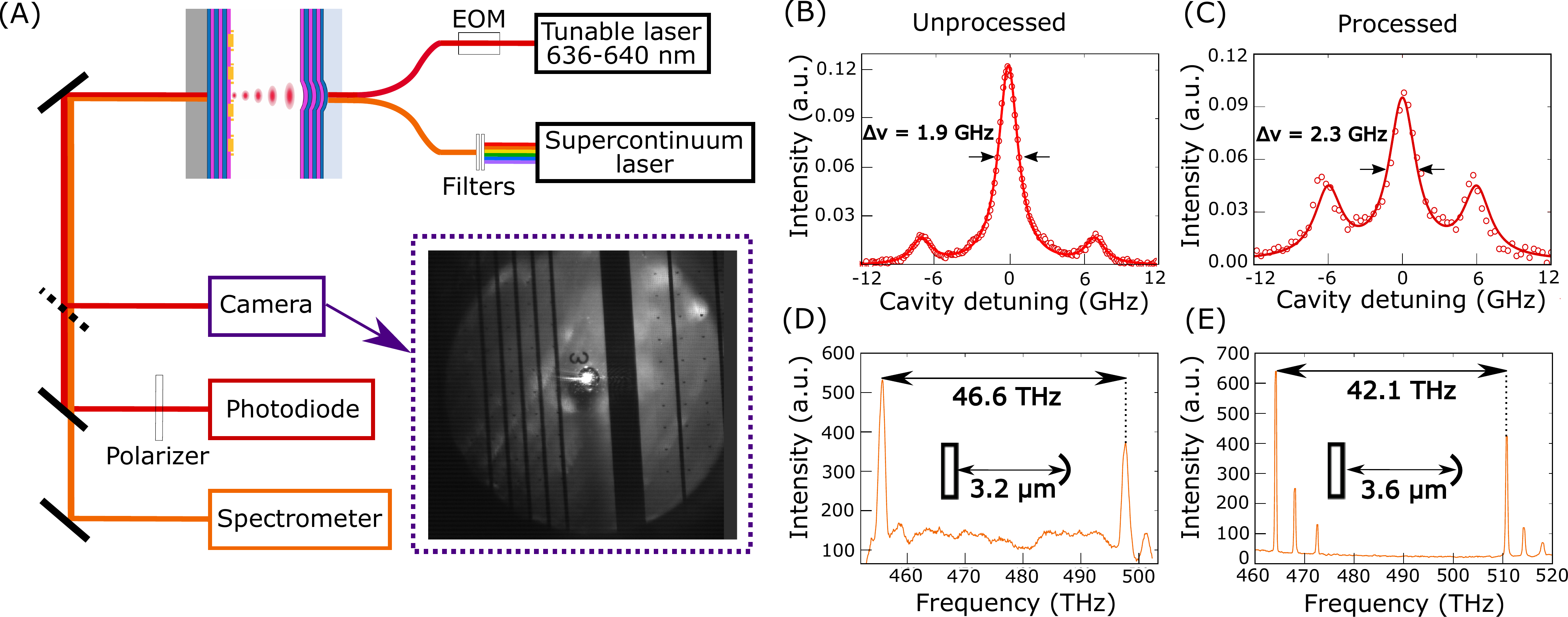}
	\caption{(Colour online) Measurement of the cavity finesse. (A) Schematic of the cavity finesse measurement setup. The cavity consists of the mirror coated ablated fiber tip glued to a custom designed fiber mount and screwed into a Newport (462-XYZ-M) stage for coarse positioning, and a planar mirror glued onto a piezo stage (PI E509.X1) for fine spectral cavity tuning. The inset shows a camera image of the patterned planar mirror and fiber microcavity on resonance. (B) Linewidth measurements of the unprocessed and (C) processed cavity are performed by exciting the cavity with a 637 nm laser (Newfocus TLB-6304) while modulating the cavity length. An EOM (Jenoptik) induces sidebands in the laser profile with 6 GHz separation that enable measuring the linewidth in frequency units. The transmitted light is collected on a photodiode (Thorlabs APD130A2) and read out on an oscilloscope. (D) Unprocessed and (E) processed cavity transmission spectra measured by coupling a supercontinuum broadband laser (Fianium SC400) into the cavity and measuring transmitted light on the spectrometer (Princeton Instruments Acton SP2500). The distance between the fundamental modes determines the cavity length (\Cref{eq:Equation_1}).\\\\}
   \label{fig:Fig_2}
\end{figure}

To verify that the presented fabrication procedure does not introduce additional losses to the planar mirror, the finesse of a cavity with a processed and unprocessed planar mirror is compared while keeping the fiber mirror unaltered. The finesse $F$ is calculated from the cavity linewidth $\delta\nu$ and free spectral range (FSR) $\nu_{FSR}$:
\begin{equation}\label{eq:Equation_1}
F = \frac{\nu_{FSR}}{\delta\nu} = \frac{c}{2\cdot L_{cav}\cdot \delta\nu},
\end{equation}
where $c$ is the speed of light and $L_{cav}$ the cavity length.

The cavity linewidth is obtained by exciting the cavity with a 637 nm laser through the fiber port while modulating the planar mirror position with a 40 Hz sinusoidal signal scanning the cavity across the resonance. The measurement setup is presented in \Cref{fig:Fig_2}(A). Before entering the cavity, the light passes through an electro-optic modulator (EOM) crating sidebands with a fixed frequency separation, used to convert the measured linewidth in length to frequency. The imperfect elliptical curvature in the fiber profile induces a polarization splitting of the cavity mode which is filtered using a polarizer in the detection path before the transmitted signal reaches the photodiode. An example of a measured linewidth for a cavity consisting of an unprocessed (processed) planar mirror is presented in \Cref{fig:Fig_2}(B) (\Cref{fig:Fig_2}(C)) for two different cavity lengths. The measured linewidth is an average over ten such single sweeps. It is to be noted that all cavity measurements were performed in the absence of a diamond slab, to compare only the processing effects on the mirror properties. For a study of the effects of a diamond slab incorporated into the cavity, see references [25,26].

The FSR is obtained by coupling a broadband supercontinuum laser into the cavity and measuring the transmitted signal on a spectrometer. \Cref{fig:Fig_2}(D) (\Cref{fig:Fig_2}(E)) shows the results of the FSR measurement for a cavity with an unprocessed (processed) planar mirror. Fundamental cavity modes can be seen as well as higher order modes at higher frequencies, which are confirmed by their shape in a camera image (not shown).
Using the measured linewidth and FSR, the finesse value of the unprocessed mirror cavity is calculated (\Cref{eq:Equation_1}) to be $F=(23\pm3)\cdot10^3$, while the finesse of the processed mirror is $F=(20\pm2)\cdot10^3$. Measured finesse values are in good agreement with the finesse values calculated from the mirror transmission and losses.
We conclude that our fabrication procedure preserves good optical properties of the mirrors.

\paragraph{Diamond membrane preparation and bonding:}

\begin{figure}
   \includegraphics[width=1\textwidth]{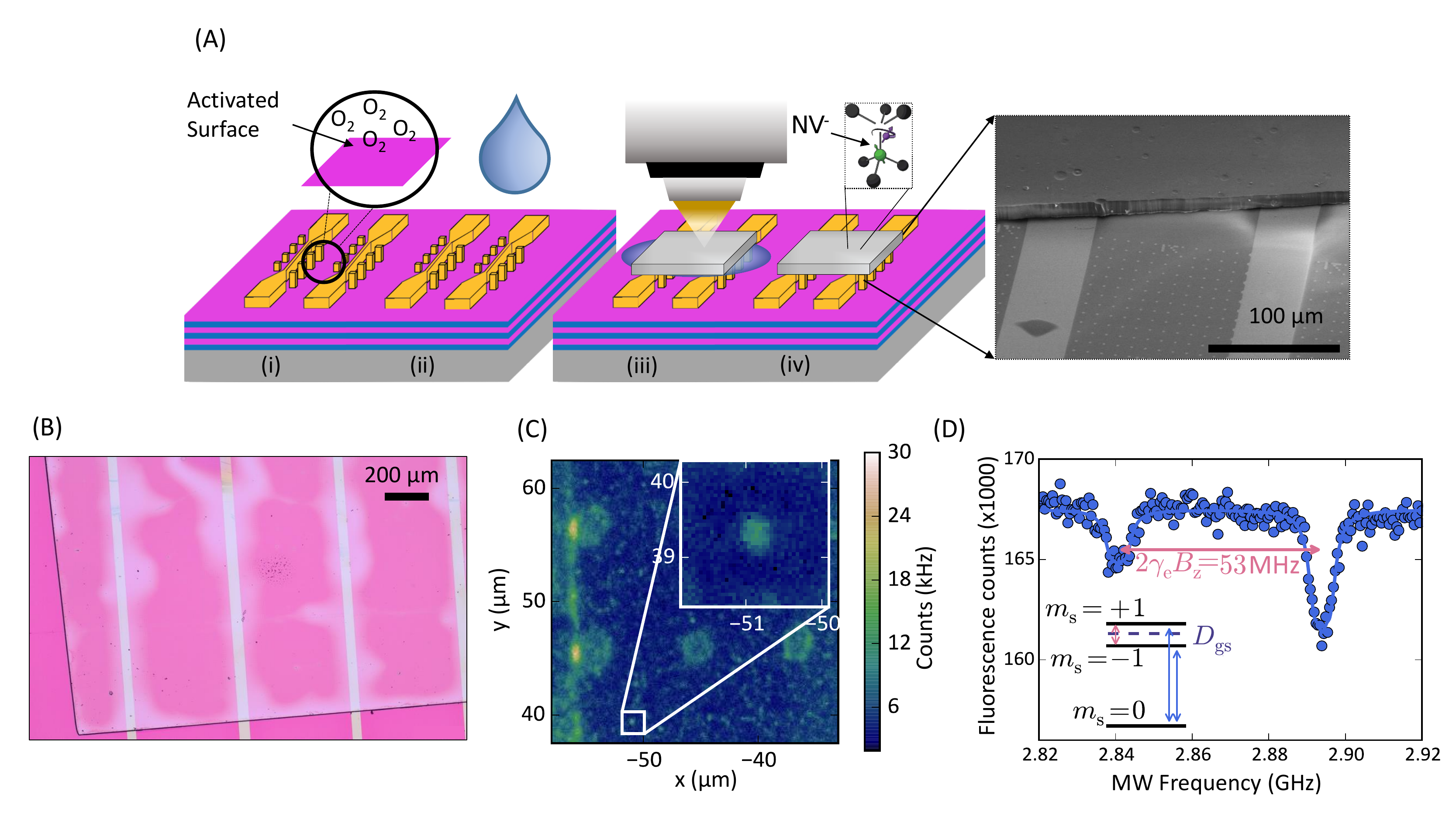}
      \caption{(Colour Online) Diamond membrane bonding and NV center spin control. (A) Diamond bonding to the patterned mirror: (i) Surface activation of the processed mirror chip with an oxygen plasma at low vacuum. (ii) Water is pipetted onto the activated surface. (iii) The diamond membrane is placed onto the patterned surface followed by drying of the water using the light from the microscope objective. (iv) The diamond membrane containing NV centers is bonded to the mirror. Inset: A scanning electron microscope (SEM) image of a bonded diamond membrane atop a patterned planar mirror. (B) Optical image of an etched 2 mm $\times$ 2 mm $\times$ 8 \SI{}{\micro m} diamond, bonded to the processed planar mirror with microwave striplines and marker array. (C) Scanning optical confocal image showing a single NV center (inset). The scans were performed at the depth of $\approx 1.5$ \SI{}{\micro m} below the surface of the diamond membrane. The 2 \SI{}{\micro m} $\times$ 2 \SI{}{\micro m} spots correspond to the photoluminescence from the square gold markers. Part of the microwave stripline is visible on the left. (D) Optically detected electron spin resonance spectrum demonstrating NV spin addressing with the embedded microwave striplines. The spectrum exhibits two resonances associated with the $m_{s} = 0 \rightarrow m_{s} = \pm 1$ spin transitions (inset), centered at the zero field splitting $D_{gs}$ = 2.87 GHz. Two resonances are separated by the Zeeman splitting 2$\gamma_{e}B_{z}$ where $\gamma_{e}$ is the NV electron spin gyromagnetic ration and $B_{z}$ is the static external magnetic field. Individual electron spin resonance dips are fitted using three Lorentzian profiles with a splitting of 2.16 MHz to account for the hyperfine splitting from the NV interaction with its host $^{14}$N nuclear spin (I$_{N}$ = 1)\citep{Neumann2009}.}
   \label{fig:Fig_3}
\end{figure}

For integration into the cavity system the large-area diamond membranes must be bonded to the processed planar mirrors. Diamond membranes are obtained by slicing and mechanically polishing 2 mm $\times$ 2 mm $\times$ 0.5 mm ${\langle100\rangle}$ bulk diamonds (Element Six), into 30 \SI{}{\micro m} thick slabs (Delaware Diamond Knives). Leftover residue and surface damage from the mechanical polishing is removed by submerging the diamond in a boiling mixture of 1:1:1 (Perchloric : Nitric : Sulfuric) acid for 1 hour, followed by the removal of several \SI{}{\micro m} on the top side of the polished diamond membrane using an Ar/Cl${_2}$ based ICP RIE (Methods). Chlorine based etching produces smooth diamond surfaces which is required to minimize scattering loss at the diamond interface within the cavity\citep{Enlund_Etch,Lee_Etch}. The final diamond roughness, measured with Atomic Force Microscopy, was found to be 0.2 nm RMS. However, prolonged exposure to chlorine etch gas has been linked to degrading optical properties of the NV centers. Introducing an O${_2}$ based plasma etching step has been found to resolve the surface chlorine contamination\citep{Chu,Riedel}.

In order to preserve optical properties of the cavity, the diamond must be fixed to the mirror without adhesives, constraining the type of bonding techniques permitted. First attempts to etch the diamond slab with the bottom surface coated with the dielectric mirror stack resulted in coating delamination, likely due to the difference in thermal expansion coefficients of the materials. Successful bonding of the diamond membrane to a processed planar mirror with an activated hydrophilic surface has been achieved via Van der Waals forces. Addition of a water droplet between the two interfaces promotes bonding via strong interfacial forces creating good optical contact between the diamond and the mirror, removing the need for adhesives.\citep{Yablonovitch1990,Pawel}.

To prepare the processed mirror for bonding, it is placed in an Oxygen plasma environment at 0.4 mbar for 45 s with 100 W, altering the hydrophobicity of the surface (\Cref{fig:Fig_3}(A,i)). Water is pipetted onto the surface of the mirror (\Cref{fig:Fig_3}(A,ii)) followed by placing the diamond membrane on top of a patterned region. Using the light from an optical microscope objective, the water is evaporated while the bonding process is monitored simultaneously (\Cref{fig:Fig_3}(A,iii)). The quality of the bond can be evaluated with visual cues as well as with a profile measurement. \Cref{fig:Fig_3}(A,iv) and \Cref{fig:Fig_3}(B) show SEM and optical images respectively of the bonded diamond. A poor bond can be identified by the appearance of Newton rings, indicative of an uneven surface and the existence of an air gap between the mirror and diamond. In the optical image, a well bonded diamond is indicated by a uniform colouration. The "milky" colouration seen near the striplines highlights the slightly elevated areas. Profilometer measurements revealed an overall height variation of 100 nm over the diamond surface bonded on top of the patterned area.

\paragraph{Electron spin resonance:}
A scanning confocal microscope is used for fluorescence imaging of the NV centers under ambient conditions with off resonant laser excitation at 532 nm. The NV centers were located in close proximity to the marker field such that they can be easily indexed and located again (\Cref{fig:Fig_3}(C)). The embedded microwave stripline, seen on the left portion of the confocal scan in \Cref{fig:Fig_3}(C), is used to identify and address the $m_{s} = 0 \rightarrow m_{s} = \pm 1$ NV center spin transitions\citep{Gruber1997,Balasubramanian2008} in the presence of a static magnetic field $B_{z} \approx 10$ G. The optically detected electron spin resonance spectrum is shown in \Cref{fig:Fig_3}(D). This demonstrates our ability to address the NV spin with microwaves when combining the stripline fabrication and diamond bonding technique.

In conclusion, we have presented the fabrication of an integrated platform for microwave control in an open, high finesse Fabry-P\'{e}rot microcavity enclosing a thin diamond membrane. Finesse measurements of the processed mirrors confirm that the fabrication procedure does not compromise their optical properties. We present a diamond bonding method that allows placing large area diamond membranes onto the cavity mirror while simultaneously enabling the microwave control of the NV center spin. The presented cavity architecture is well suited for enabling enhancement of the NV centers resonant emission along with the control of their spins.\\

The authors wish to thank P. Latawiec for helpful discussions. M.L. wishes to acknowledge the support of QuTech during his sabbatical stay. M.S.Z.L. acknowledges the Dutch Liberation Scholarship Programme. This work was supported by the Dutch Organization for Fundamental Research on Matter (FOM), Dutch Technology Foundation (STW), the Netherlands Organization for Scientific Research (NWO) through a VICI grant, the EU S3NANO program and the European Research Council through a Starting Grant.\\

Methods:\\
Markers and stripline etching procedure:
The planar mirrors are etched with an SF${_6}$ and O${_2}$ based ICP RIE technique in an (Oxford Instruments PlasmaLab 100) etcher with the following parameters: SF$_6$(Ar) gas flow - 50(10) sccm, ICP (RF) Power - 700(150) W, Base Pressure - 0.026 mBar, Temperature - ${50^\circ}$C, Time - 26 s followed by O${_2}$ gas flow - 50 sccm, ICP(RF) Power - 750(20) W, Base Pressure - 0.030 mBar, Temperature - ${50^\circ}$C, Time - 8 s at a combined rate of 1.9 nm/s, removing 65 nm of mirror material. \\
The diamond etch parameters are: Ar(Cl${_2}$) gas flow-10(20) sccm, ICP (RF) Power - 500(200) W, Base Pressure - 0.01 mBar, Temperature - ${30^\circ}$C with a resulting etch rate of 2.5 \SI{}{\micro m}/hr.

\bibliography{Bogdanovic_Liddy_bib}
\end{document}